\documentclass[11pt,a4paper]{article}

\usepackage{amsmath,amssymb}
\usepackage{lipsum}
\usepackage{enumitem}
\usepackage{soul}
\usepackage{graphicx}
\usepackage{color}
\usepackage{slashed}
\usepackage{bbold}
\usepackage{wasysym}
\usepackage{graphicx}

\usepackage{multirow}

\usepackage[
linktocpage=true,
  colorlinks   = true,    
  urlcolor     = blue,    
  linkcolor    = blue,    
  citecolor    = blue      
]{hyperref}
\usepackage[sort&compress,square,numbers]{natbib}

\newcommand{\TeV}{{\ensuremath\rm TeV}}
\newcommand{\GeV}{{\ensuremath\rm GeV}}

\newcommand{\lb}{\left(}
\newcommand{\rb}{\right)}
\newcommand{\fb}{\ensuremath\rm fb}

\newcommand{\HSv}[1]{\texttt{HiggsSignals-#1}}
\newcommand{\HBv}[1]{\texttt{HiggsBounds-#1}}
\newcommand{\HS}{\texttt{HiggsSignals}}
\newcommand{\HB}{\texttt{HiggsBounds}}
\newcommand{\eqn}{equation}
\newcommand{\lam}{\lambda}

\def\mev{\;\hbox{MeV}}
\def\gev{\;\hbox{GeV}}

\begin{document}
\begin{center}
{\LARGE \textbf{Benchmarking the Inert Doublet Model\\ for  $e^+e^-$ colliders}}
\\ [1cm]
{\large {Jan Kalinowski,$^{a}$ Wojciech Kotlarski,$^{b}$ Tania Robens,$^{c,d}$ Dorota Soko\l owska$^{a,e}$\\ and Aleksander Filip \.Zarnecki$^{a}$
}}
\\[1cm]
\end{center}

\hspace{1cm}
\begin{minipage}{14cm}
\textit{ 
\noindent $^a$ Faculty of Physics, University of Warsaw, 
ul.~Pasteura 5, 02--093 Warsaw, Poland\\[0.1cm]
\noindent $^b$ 
Institut f\"ur Kern- und Teilchenphysik, TU Dresden, 
01069 Dresden, Germany\\[0.1cm]
$^c$ MTA-DE Particle Physics Research Group, University of Debrecen, 
4010 Debrecen, Hungary\\[0.1cm]
$^d$ Theoretical Physics Division, Rudjer Boskovic Institute,
10002 Zagreb, Croatia\\[0.1cm]
$^e$ International Institute of Physics, Universidade Federal do Rio Grande do Norte,
Campus Universitario, Lagoa Nova, Natal-RN 59078-970, Brazil
}
\end{minipage}

\vspace{0.7cm}

\begin{center}
\textbf{Abstract}

\begin{quote}
We present benchmarks for the Inert Doublet Model, a Two Higgs Doublet Model with a dark matter candidate. They are consistent with current constraints on direct detection, including the most recent bounds from the XENON1T experiment and relic density of dark matter, as well as with {known} collider and low-energy limits. We focus on parameter choices that promise detectable signals at lepton colliders via pair-production of $H^+H^-$ and $HA$. For these we choose a large variety of benchmark points with different kinematic features, leading to distinctly different final states in order to cover the large variety of collider signatures that can result from the model.

\end{quote}

\end{center}
\vspace{10mm}

\def\thefootnote{\arabic{footnote}}
\setcounter{footnote}{0}

\section{Introduction}
The results from the ATLAS and CMS collaborations from Run I and ongoing Run II are in good agreement with the predictions of the Standard Model (SM) \cite{ICHEP:ATLAS,ICHEP:CMS}. Although the discovered Higgs particle appears to be consistent with the expectations for a SM Higgs boson, both the experimental uncertainties and theoretical speculations still leave room for new physics.
In particular the scalar sector can provide intriguing scenarios in this respect which should be further scrutinized.  

Although experimental collider data is in good agreement with predictions of the Standard Model alone, 
{a number of} non-collider observations can only be described in models containing additional (new physics) constituents. A prime example for this is dark matter (DM).
Within the standard model of cosmology, 
the Planck mission data \cite{Ade:2015xua} implies that nearly 85\% of the total matter
content in the universe is dark. 
However, so far only the gravitational interactions of these hypothetical particles have been detected, 
and a fundamental nature
of DM remains largely unknown. Since the Standard Model of
elementary particles does not contain a viable DM candidate, any
evidence of DM in the direct detection or  indirect detection
experiments  or production at colliders would be a signal of new
physics, the discovery of which is arguably one of the most important
goals in the field. 

An intriguing extension of the SM scalar sector is the Inert Doublet Model (IDM) which features a dark matter candidate \cite{Deshpande:1977rw,Cao:2007rm,Barbieri:2006dq}. In this two Higgs doublet model a discrete $Z_2$ symmetry (called $D$-symmetry) is imposed, with the following transformation properties:
\begin{equation}
\phi_S\to \phi_S, \,\, \phi_D \to - \phi_D, \,\,
\text{SM} \to \text{SM},
\end{equation}
where the $\phi_S$ doublet plays the same role as the corresponding  doublet in the SM, {providing} the SM-like Higgs particle. This doublet is even under the $D$-symmetry, while the second doublet, the inert (or dark) $\phi_D$, is $D$-odd and contains four scalars, two charged and two neutral ones, labelled $H^\pm$ and $H,A$, respectively.
The above symmetry renders the additional SU(2)$_L$ doublet $\phi_D$ inert, $i.e.$ prevents its couplings to  the SM matter sector, thereby providing a dark matter candidate. In the rest of this work, we consider cases where $H$ serves as the dark matter candidate of the model.

The IDM was first discussed in \cite{Deshpande:1977rw} and later in \cite{Cao:2007rm,Barbieri:2006dq}.  The model was further studied in \cite{LopezHonorez:2006gr,Honorez:2010re,Dolle:2009fn,Goudelis:2013uca,Krawczyk:2013jta}, followed with further analyses of the IDM at colliders \cite{Lundstrom:2008ai,Dolle:2009ft,Gustafsson:2012aj,Aoki:2013lhm,Ho:2013spa,Arhrib:2013ela,Arhrib:2012ia,Swiezewska:2012eh,Krawczyk:2013jta,Ginzburg:2014ora,Belanger:2015kga,Blinov:2015qva,Ilnicka:2015jba,Hashemi:2015swh,Poulose:2016lvz,Datta:2016nfz,Kanemura:2016sos,Akeroyd:2016ymd,Wan:2018eaz,Ilnicka:2018def,Belyaev:2018ext}. 

We here present a set of the IDM benchmarks proposed for detailed studies at the future $e^+e^-$ colliders ILC and CLIC.
They have been selected from  updates of the scan presented in \cite{Ilnicka:2015jba,Ilnicka:2018def} and represent
distinct features for two prominent production processes at linear colliders, $e^{+}e^{-} \rightarrow
  H^{+}H^{-}$ and $e^{+}e^{-}\rightarrow AH$.  Our benchmarks are designed to cover all interesting parameter space, featuring different mass splittings between $H$ and other dark particles, leading to distinct collider signatures.

  The outline of this paper is as follows. We start with a short description of the IDM in section \ref{sec:model}, followed with a discussion of  {theoretical constraints and} current experimental limits in section \ref{sec:constraints}. In section \ref{sec:benchmarks} we define benchmark points that pass all constraints and discuss their applications to further studies  {at the first stage of linear colliders running at 250, 380 and 500 GeV center-of-mass energies. Benchmarks suitable for testing at high energy stages of 1, 1.5 and 3 TeV linear colliders are presented in section \ref{sec:high}. We also comment on the impact of future XENON-nT measurements and prospects of testing the IDM at the LHC in section \ref{sec:lhc}.}

\section{The IDM \label{sec:model}}

The scalar sector of the IDM consists of two {SU(2)$_{L}$} doublets of complex scalar fields,   $\phi_S$  and $\phi_D$, {{with the $ D$-symmetric potential:}}
\begin{equation}\begin{array}{c}
V=-\frac{1}{2}\left[m_{11}^2(\phi_S^\dagger\phi_S)\!+\! m_{22}^2(\phi_D^\dagger\phi_D)\right]+
\frac{\lambda_1}{2}(\phi_S^\dagger\phi_S)^2\! 
+\!\frac{\lambda_2}{2}(\phi_D^\dagger\phi_D)^2\\[2mm]+\!\lambda_3(\phi_S^\dagger\phi_S)(\phi_D^\dagger\phi_D)\!
\!+\!\lambda_4(\phi_S^\dagger\phi_D)(\phi_D^\dagger\phi_S) +\frac{\lambda_5}{2}\left[(\phi_S^\dagger\phi_D)^2\!
+\!(\phi_D^\dagger\phi_S)^2\right].
\end{array}\label{pot}\end{equation}
Exact $D$-symmetry implies that only  $\phi_S$ can acquire  {a} nonzero vacuum expectation value ($v$). As a result the  scalar field{s}  {in $\phi_D$} do not mix with the SM-like field {from} $\phi_S$, and {the} lightest particle of  {the} dark sector is stable.  
The dark sector  {contains} four new particles: $H$, $A$ and $H^{\pm}$.  {We here choose $H$ to denote the} dark matter candidate.  {{A priori, any of the new scalars can function as a dark matter candidate. However, we neglect the choice of a charged dark matter candidate, as these are strongly constrained \cite{Chuzhoy:2008zy}.
As inert scalars do not couple to fermions, it is not possible to assign a definite CP-property to them. Therefore, choosing $A$ instead of $H$ to be the lightest particle changes only the meaning of $\lam_5$, with rephasing of $\lambda_5 \to - \lambda_5$, but not the overall phenomenology of the model, cf. Appendix A in \cite{Ilnicka:2015jba}.}}  \\

 After electroweak symmetry breaking, the model contains seven free parameters. Agreement with the Higgs boson discovery and electroweak precision observables fixes the SM-like Higgs mass  {$M_h$} and $v$,  {and we are left with} five free parameters, which we take as
\begin{\eqn}\label{eq:physbas}
M_H, M_A, M_{H^{\pm}}, \lam_2, \lam_{345},
\end{\eqn}
 where  {the} $\lambda$'s refer to coupling{s} within the dark sector and to the SM-like Higgs respectively. In the following, we will use the abbreviation  $\lam_{345} = \lam_{3}+\lam_{4}+\lam_{5}$.

\section{Experimental and theoretical constraints \label{sec:constraints}}

In this work, we make use of the tool chain used in \cite{Ilnicka:2015jba}, and explicitly follow the scan procedure described therein; however, we update several constraints as discussed below. Explicit benchmark points (BPs) for the IDM, taking all constraints viable at that time into account, have been presented in \cite{Ilnicka:2015jba,deFlorian:2016spz}, which focus on processes at the LHC; some of these were already investigated in a  linear collider context in \cite{Hashemi:2015swh}.
Ref. \cite{Ilnicka:2018def} shows how the available parameter space for certain scenarios is further limited by more recent constraints.

As the experimental constraints have evolved significantly since that
time, we decided to define the new set of benchmark points, fulfilling
the updated constraints and focused on the detailed analysis of $e^+
e^-$ collider sensitivity. Unless stated otherwise, all considered BPs 
fulfil the latest experimental limits; 
following \cite{Ilnicka:2015jba} (see also the discussion in \cite{Belyaev:2016lok}), we do not require the IDM to provide $100\%$ of the dark matter relic density.
Below we briefly summarize the imposed constraints, emphasizing on updates with respect to \cite{Ilnicka:2015jba}, and
describe the set-up to find good BPs and discuss the obtained limits.

\subsection{Theoretical and experimental constraints}

\textbf{Positivity constraints}: we require that the potential is bounded from
 below, therefore no field configuration leads to $V\,\rightarrow\,-\infty$,
  resulting in tree-level relations \cite{Nie:1998yn}
\begin{equation}
\lam_1\,>\,0,\,\lam_2\,>\,0,\,\lam_3+\sqrt{\lam_1 \lam_2}>0,\,\lam_{345}+\sqrt{\lam_1 \lam_2}\,>0.
\end{equation}
These relations hold on tree level and in this work we do not consider higher-order contributions, which in principle could lead to change in stability of
  the electroweak vacuum \cite{Goudelis:2013uca, Swiezewska:2015paa}.

\textbf{Perturbative unitarity}: we require the scalar {$2\,\rightarrow\,2$}
 scattering matrix to be unitary, i.e. all eigenvalues of scattering
  matrices for scalars  with specific hypercharge and isospin should 
  satisfy $|L_i|\,\leq\,16\,\pi$ \cite{Chanowitz:1985hj,Ginzburg:2005dt}. Furthermore, 
  we require all quartic scalar couplings to be perturbative, i.e. to 
  take absolute values $\leq\,4\,\pi$. 

\textbf{Global minimum}: in the IDM two neutral minima can coexist 
even at tree level. Unless the following relation is satisfied
\begin{equation}\label{eq:invac}
\frac{m_{11}^2}{\sqrt{\lam_1}}\,\geq\,\frac{m_{22}^2}{\sqrt{\lam_2}},
\end{equation}
the inert minimum is only a local one, with the global vacuum 
corresponding to the case of massless fermions \cite{Ginzburg:2010wa}.
We impose the above relation in our scan.

\textbf{Higgs mass and signal strengths}: the mass of the SM-like 
Higgs boson $h$ is set to $$M_h=125.1\,\gev \label{Eq:mhexp},$$ in 
agreement with limits from ATLAS and CMS experiments \cite{Aad:2015zhl}, 
while the total width of the SM-like Higgs boson obeys an upper
 limit of  {\cite{CMS:2018bwq}}  
\begin{\eqn}\label{eq:gtot}  
 {\Gamma_\text{tot}\,\leq\,9\,\mev.}
\end{\eqn}

In the IDM, the total width of the SM-like state can obtain modifications from the following two contributions. For dark matter masses $M_H\,\leq\,M_h/2$, invisible decays of the 125 \GeV~resonance can lead to large additional contributions. Therefore, in these scenarios the above bound poses one of the most dominant constraints, especially affecting $\lam_{345}$ \cite{Ilnicka:2015jba,Ilnicka:2018def}. Furthermore, the partial decay width of $h$ to diphoton final states can be altered significantly \cite{Swiezewska:2012eh,Krawczyk:2013jta}, as the new physics corrections are formally of the same order as the SM process. This leads to a clear distinction between allowed and forbidden regions in the $\lb\lam_{345}, M_{H^\pm}\rb$ plane \cite{Ilnicka:2015jba,Ilnicka:2018def}. The Run I combined ATLAS and CMS limit for $h\to \gamma \gamma$ signal strength is given by $\mu_{\gamma \gamma} = 1.14^{+0.38}_{-0.36}$ \cite{Khachatryan:2016vau}. In our analysis we use both the upper limit (\ref{eq:gtot}) and require agreement within $2\,\sigma$ for the prediction of $h\,\rightarrow\,\gamma\,\gamma$. We furthermore check agreement with all other branching ratios of the 125 Higgs on the $2\,\sigma$ level using the publicly available tool {\HSv2.2.1beta}~\cite{Bechtle:2013xfa}, {and require $\Delta \chi^2\,\leq\,11.31$}, corresponding to a $95 \%$ confidence level. 

\textbf{Gauge bosons width}: introduction of light new particles could in principle significantly change the total width of electroweak gauge bosons {(cf. e.g. \cite{Patrignani:2016xqp})}. To ensure that $W^\pm \to H H^\pm$ and $Z\to HA,H^+H^-$ decay channels are kinematically forbidden we set:
\begin{equation}\label{eq:gwgz}
M_{A,H}+M_{H^\pm}\,\geq\,M_W,\,M_A+M_H\,\geq\,M_Z,\,2\,M_{H^\pm}\,\geq\,M_Z.
\end{equation} 

\textbf{Electroweak precision tests (EWPT)}: we call for a $2\,\sigma$ (i.e. $95 \%$ C.L.) agreement with electroweak precision observables, parametrized through the electroweak oblique parameters $S,T,U$ \cite{Altarelli:1990zd,Peskin:1990zt,Peskin:1991sw,Maksymyk:1993zm}. In our work, calculations were done through the routine implemented in the Two Higgs Doublet Model Calculator (\texttt{2HDMC}) tool \cite{Eriksson:2009ws}, which checks whenever model predictions fall within the observed parameter range \cite{Baak:2014ora}.

\textbf{Charged scalar mass and lifetime}: we take a conservative lower estimate on the mass of $M_{H^\pm}$ following analysis in \cite{Pierce:2007ut} to be 
\begin{\eqn}\label{eq:mhplow}
M_{H^\pm}\,\geq\,70\,\gev.
\end{\eqn}
We also set an upper limit on the charged scalar lifetime of 

\begin{\eqn}\label{eq:limcharged}
\tau\,\leq\,10^{-7}\,s,
\end{\eqn}
in order to evade bounds from quasi-stable charged particle searches. This translates to a lower bound on the total decay width of the charged scalar $H^\pm$ of $\Gamma_\text{tot}\,\geq\,6.58\,\times\,10^{-18}\,\gev$. In \cite{Heisig:2018kfq}, the authors show that limits from a recast of heavy charged particle searches at the LHC \cite{Khachatryan:2015lla} can set bounds on the lifetimes of $H^\pm$ between $10^{-9}$ and $10^{-5}\,s$, with the lower value prevailing for charged masses $M_{H^\pm}\,\lesssim\,500\,\GeV$. For most of the parameter space, the limit (\ref{eq:limcharged}) therefore corresponds to a conservative upper limit.

\textbf{Collider searches for new physics}: we require agreement with the  null-searches from the LEP, Tevatron, and LHC experiments. We use the publicly available tool \HBv5.2.0beta \cite{Bechtle:2008jh, Bechtle:2011sb, Bechtle:2013wla,Bechtle:2015pma}. In addition
the reinterpreted LEP II searches for supersymmetric particles analysis exclude the region of masses in the IDM where simultaneously \cite{Lundstrom:2008ai}
\begin{equation}\label{eq:leprec}
M_A\,\leq\,100\,\gev,\,M_H\,\leq\,80\,\gev,\,\, \Delta M {(A,H)}\,\geq\,8\,\gev,
\end{equation}
as it would lead to a visible di-jet or di-lepton signal. After taking into account all the above limits we are outside of the region excluded due to the reinterpretation of the supersymmetry analysis from LHC Run I \cite{Belanger:2015kga}.

\textbf{Dark matter phenomenology}: we apply dark matter relic density limits obtained by the Planck experiment \cite{Ade:2015xua}:
\begin{equation}\label{eq:planck}
\Omega_c\,h^2\,=\,0.1197\,\pm\,0.0022.
\end{equation}
For a DM candidate that provides 100\% of observed DM in the Universe we require the above bound to be fulfilled within the 2$\sigma$ limit. However, we also allow for the case where $H$ is only a subdominant DM candidate, with 
\begin{\eqn}\label{eq:om_limapp}
\Omega_H h^2 < \Omega_c\,h^2
\end{\eqn}
 (see \cite{Ilnicka:2015jba,Belyaev:2016lok,Ilnicka:2018def}). In such a scenario, additional dark matter candidates would be needed in order to account for the missing relic density.
In the results presented here, we apply XENON1T limits \cite{Aprile:2018dbl}\footnote{We use a digitized format of that data available from \cite{PhenoData}.}. These supersede previous bounds applied e.g. in \cite{Ilnicka:2018def} in relevant regions of parameter space, and are therefore crucial for a correct determination of the available parameter space, especially for low dark masses.

Results from indirect detection experiments, e.g. Fermi-LAT \cite{Fermi-LAT:2016uux}, give less stringent constraints than collider and direct detection experiments discussed above \cite{Goudelis:2013uca,Belyaev:2016lok}. A number of planned DM indirect detection experiments, mainly the Cherenkov Telescope Array, will be able to probe the heavy mass region, with DM particle heavier than 500 GeV \cite{Queiroz:2015utg}. Furthermore, if the reported gamma-ray excess from the Galactic center is of DM origin \cite{TheFermi-LAT:2017vmf}, it can be explained by the IDM with dark matter masses near the Higgs resonance or around 72 GeV \cite{Eiteneuer:2017hoh}. All dark matter variables were calculated with the use of \texttt{micrOmegas} version 4.3.5 \cite{Barducci:2016pcb}.

\subsection{Scan setup and limits}

After fixing the value of $M_h$, and hence both $\lambda_1$ and $m_{11}^2$, we are left with five {independent} input parameters for the scan: three masses of dark scalars $M_{A,H,H^\pm}$ and two couplings, $\lambda_{345},\lambda_2$. {In the initial setup of our scan,} masses take values between 0 and 1 TeV, with $M_H$ always being the lightest and $M_{H^\pm} \geq 70$ GeV. Unless stated otherwise, scalar couplings fall in the range of $\lambda_2 \in [0;4.5],\lambda_{345}\in [-1.5;4\pi]$. 

In order to get interesting benchmark points we follow the procedure described in \cite{Ilnicka:2015jba}. All constraints described in the previous section are checked in steps, with the aid of publicly available tools. In the process we track the impact of each exclusion criterion.

The first step contains a check through the \texttt{2HDMC} in order to establish agreement with theoretical constraints (positivity, stability, perturbative unitarity). The same code is used to check the SM-like Higgs and electroweak gauge bosons widths, the decay rates of $h\,\rightarrow\,\text{invisible}$, and $h\,\rightarrow\,\gamma\gamma$, properties of charged scalar (lifetime, mass) as well as the EWPT observables. Points that have passed the first step are then checked against limits from collider searches, in particular the Higgs signal strength limits, with the use of \HB\,  and \HS. Points which passed the collider test are then confronted with limits from DM phenomenology, i.e. the relic density constraints with upper Planck bound (points that correspond to 100\% of measured $\Omega_c h^2$ are selected at later stage) and direct detection limits from XENON1T, with the use of \texttt{micrOmegas}.

Although the IDM is one of the simplest extensions of the SM and has a limited number of parameters, it is still difficult to establish which constraint has the greatest impact on excluding a given point in parameter space. In the following, we point to some generic features that however can be used for a clear distinction in the model parameters space (see also \cite{Ilnicka:2015jba}). 

It is important to emphasize that the couplings that govern the production and decay processes at $e^+e^-$ colliders are mainly determined by the electroweak parameters of the SM; the additional parameters in the potential do not play any significant role for the allowed scenarios. On the other side, the parameter $\lam_{345}$ is especially sensitive to constraints from dark matter observations. This nicely demonstrates the important complementarity of collider and astrophysical measurements in constraining the IDM parameter space. 

We divide the discussion in two different subsections
\begin{itemize}
\item[$(i)$] constraints on the masses of the dark scalars, with the constraint given in eqn.~(\ref{eq:om_limapp}) for the dark matter relic density;
\item[$(ii)$] additional limits we obtain when requiring exact relic density.
\end{itemize}
We only briefly discuss the second point here and refer to the literature (see e.g. \cite{Belyaev:2016lok}) for further details.

\subsubsection{Limits on masses}
The collider phenomenology of the IDM is mainly determined by the dark scalar masses, as all relevant production and decay channels are governed by electroweak couplings and the corresponding SM parameters. As dominant constraints for the regions for $M_H\,\leq\,M_h/2$ and $M_H\,\geq\,M_h/2$ originate from different sources, we will discuss these separately.

\begin{itemize}
\item{}In general, constraints arising from EWPT are of a great importance to our studies. As found in \cite{Ilnicka:2015jba,Belyaev:2016lok,LopezHonorez:2010tb}, mass splittings between inert particles are heavily limited by the S,T,U parameters. 
First, only moderate mass splittings are allowed by EWPT data, with the preferred value of $M_{H^\pm} - M_A$ below 100 GeV. Also, there is a hierarchy between masses, with the charged scalar being the heaviest particle. The reverse relation is not excluded, however it will lead to a larger tension with combined S and T limits.  In general, points with moderate mass splittings are preferred, especially for dark matter candidates with masses $\geq\,300\,\GeV$, where splittings between $M_A$ and $M_H$ are typically of order $10\,\%$ or lower. For masses $M_H\,\leq\,100\,\GeV$, we found that relatively large mass differences are allowed between the two neutral scalars. The mass hierarchy and constraints on mass splittings can influence cascade decays of inert particles.
\item In addition to eqn.~(\ref{eq:mhplow}), the measurement of $h\,\rightarrow\,\gamma\gamma$ puts a lower limit on the charged Higgs mass as a function of $\lam_{345}$, cf. e.g. \cite{Ilnicka:2018def}. 
\item masses of DM particles below 45 GeV are excluded \cite{Ilnicka:2015jba,Belyaev:2016lok}. For $M_H\,\leq\,M_h/2$ limits from the invisible branching ratio of the SM Higgs lead to relatively low values of $\lam_{345}$, which in turn result in  the relic density exceeding the measured value by orders of magnitude or the mass splittings changing electroweak gauge bosons widths\footnote{See e.g. detailed discussions in \cite{Belyaev:2016lok}.} significantly. 
\item{} for dark scalars with masses $\lesssim\,100\,\GeV$, additional specific constraints are given by eqn.~(\ref{eq:leprec}).
\end{itemize} 

The parameters $\lam_2$ and $\lam_{345}$ are also constrained from combined positivity, unitarity and global minimum conditions \cite{Swiezewska:2012ej}. The exact value of $\lam_2$ parameter would matter in studies that consider interaction between inert particles, for example for the astrophysical implications of self-interacting dark matter, or if the loop processes like $HH \to \gamma \gamma$ through the $H^+H^-$ loop are considered, this is however beyond the scope of this work. 

Dark matter constraints give major bounds on allowed values of $\lam_{345}$. In general, $|\lam_{345}|\,\lesssim\, 1.5$ for dark masses up to 1 \TeV. For $M_H\,\lesssim\,M_h/2$, this limit decreases to $|\lam_{345}|\,\lesssim\,0.006$, due to the inclusion of results presented in \cite{Aprile:2018dbl}. Limits from Higgs signal strength measurements also limit this parameter, albeit less strongly than dark matter constraints. If we allow for a subdominant dark matter candidate, the parameter space largely opens up for regions where the dark matter relic density is much lower than the Planck value. Direct detection limits are then rescaled and therefore considerably relaxed (see also \cite{Ilnicka:2015jba}). A prominent example for this is the region where $M_H\,\sim\,M_h/2$, leading to large annihilation cross sections. This broadens especially the allowed range for $\lam_{345}$.

In summary, $M_{H^\pm}$ and $M_A$ are relatively degenerate throughout the allowed parameter space. The splittings between these and the dark matter candidate depend on the DM mass and can become quite large, especially for low $M_{H}$. The couplings $\lam_2$ and $\lam_{345}$ do not play a significant role in collider phenomenology.

\subsubsection{Requiring exact relic density}
\label{sec:exact}

Requiring the model to render exact relic density, as specified by eqn.~(\ref{eq:planck}), puts additional constraints on the model. These are mainly on the coupling combination $\lam_{345}$, but also affect the values of possible dark scalar masses in such a scenario.

\begin{itemize}
\item{}If $H$ is the only source of DM in the Universe then it is possible to find good points  {only for masses 55\,GeV $\lesssim M_H \lesssim$ 75\,GeV, where the allowed range of dark scalar masses is determined by electroweak constraints, without additional fine-tuning.}
\item{} {In the region $75\,\GeV \lesssim M_H \lesssim 500\,\GeV$ DM direct detection searches exclude all points that have exact relic density. Points that have passed all other constraints provide only a subdominant dark matter candidate.}  
\item{}For $M_{H}$ larger than 500\,GeV extreme fine-tuning of the dark masses is required in order to obtain exact relic density; cf. the discussion in \cite{Belyaev:2016lok}. The authors of that reference find that $\mathcal{O}\lb \GeV\rb$ mass splittings between the dark scalars would be required in order to obtain the correct relic density; in turn, this can lead to long-lived particles at the LHC, {where minimal allowed mass splittings are $\mathcal{O}\lb 0.2\,\GeV \rb$ \cite{Heisig:2018kfq}}. In our work, we do not study this particularly fine-tuned region in more detail.
\end{itemize}

\section{Benchmark Points \label{sec:benchmarks}}

As the aim of this note is to present benchmarks useful for further studies at linear colliders, we are particularly interested in points that provide an observable signal.
For the possible signal we take the pair-production processes
\begin{eqnarray}
  e^{+}e^{-} \to  H^{+}H^{-} & \text{and}&   e^{+}e^{-}  \to   AH \label{processes}
  \end{eqnarray}
for charged and  neutral scalar production. 
The s-channel production $e^+e^- \to h \to AA$ is also possible, however it is suppressed by a small electron Yukawa coupling. The dark scalars $H^\pm$ and $A$ will then decay into a virtual or on-shell electroweak gauge boson and the dark matter candidate. Here, we first concentrate on center-of-mass energies of $\sqrt{s}\,\in\left\{ 250; 380; 500 \right\} \GeV$, therefore constraining ourselves to scenarios where either $M_{H^+}\,\leq\,250\,\GeV$ or $M_H+M_A\,\leq\,500\,\GeV$ (or both). In section \ref{sec:high}, we extend the analysis to contain higher dark scalar masses, in order to investigate the collider reach of the high energy ILC at 1\,TeV and higher energy stages of CLIC at 1.5\,TeV and 3\,TeV \cite{CLIC:2016zwp}.
In this work, we consider benchmarks with both on-shell and off-shell intermediate gauge bosons, as these differ in the collider phenomenology, and different cut strategies have to be applied.

Our benchmark selection was done in the following steps:
\begin{enumerate}
\item{}Benchmark candidates were generated by employing the scan presented in \cite{Ilnicka:2015jba}, with updated experimental constraints as discussed above. We selected about 7500 benchmark candidates which fulfill all constraints (we allow for under-abundant dark matter, therefore relaxing bounds from direct detection considerably in certain regions of parameter space).
\item{}We then calculated the production cross sections at $250,\,380\,$ and $500\,\GeV$ center-of-mass energies for processes~(\ref{processes}); we required a minimal production cross section of $10\,\fb$ to classify a point as "accessible" for a certain process/energy stage;
\item{}if possible, we required a high-enough relic density, providing at least $50\%$ of the value observed by the Planck collaboration;
\item{}finally we selected benchmark points corresponding to different accessibility at the subsequent energy stages and different kinematical configurations, namely on-shell vs off-shell intermediate gauge bosons.
\end{enumerate}

The calculation of cross sections was performed using input files generated by {\tt SARAH 4.13.0} \cite{Staub:2012pb,Staub:2013tta,Staub:2015kfa} and {\tt SPheno 4.0.3} \cite{Porod:2003um,Porod:2011nf}, which were passed to {\tt WHizard 2.2.8}~\cite{Moretti:2001zz,Kilian:2007gr}. Initial state radiation was taken into account but not the beam luminosity spectra.

Point 2 above allows us to categorize different classes of benchmarks, which we label "XYZ" in the following, with  X=1(2) corresponding to  1 (2) production processes accessible at 250\,GeV, while Y and Z are defined accordingly for 380\,GeV and 500\,GeV. Within a certain category, we then considered different mass splitting configurations, in order to cover all possible typical parameter configurations leading to distinct collider signatures that can be generated by the IDM. 
The final selection of benchmark points, focusing on kinematic properties and selection criteria, is given in table \ref{tab:bench}. The table also contains the complete set of independent parameters for each point, as well as the relic density.

\begin{table}[p]
\begin{center}
  \small
  \begin{tabular}{|l|l|l|l|c|c|c|l|l|l|}
\hline
\multirow{2}{*}{No.} & \multirow{2}{*}{$M_H$} & \multirow{2}{*}{$M_A$} & \multirow{2}{*}{$M_{H^\pm}$} & $Z$ & $W$ & DM & \multirow{2}{*}{$\lambda_2$} & \multirow{2}{*}{$\lambda_{345}$} & \multirow{2}{*}{$\Omega_H h^2$}\\[-1mm] 
 & & & & on-shell &  on-shell & $>$50\%  &  &  & \\ \hline
\multicolumn{10}{|l|}{222}\\ \hline
\textbf{BP1} & 72.77 & 107.803 & 114.639 & &&$\checked$& 1.44513 & -0.00440723 & 0.12007\\\hline
BP2 & 65 & 71.525 & 112.85 & &&$\checked$ & 0.779115 & 0.0004 & 0.070807\\ \hline
BP3 & 67.07 & 73.222 & 96.73 &&&$\checked$& 0 & 0.00738 & 0.061622\\ \hline
\multicolumn{10}{|l|}{122}\\ \hline
BP4 & 73.68 & 100.112 & 145.728 & &&$\checked$& 2.08602 & -0.00440723 & 0.089249\\ \hline
\textbf{BP6} & 72.14 & 109.548 & 154.761 & &$\checked$&$\checked$& 0.0125664 & -0.00234 & 0.11708\\ \hline
\multicolumn{10}{|l|}{112}\\ \hline
BP7 & 76.55 & 134.563 & 174.367 & &$\checked$&& 1.94779 & 0.0044 & 0.031402\\ \hline
\textbf{BP8} & 70.91 & 148.664 & 175.89 & &$\checked$&$\checked$& 0.439823 & 0.0051 & 0.124\\ \hline
BP9 & 56.78 & 166.22 & 178.24 & $\checked$&$\checked$&$\checked$& 0.502655 & 0.00338 & 0.081268\\ \hline
BP23 & 62.69 & 162.397 & 190.822 &  $\checked$&$\checked$&$\checked$& 2.63894 & 0.0056 & 0.064038\\ \hline
\multicolumn{10}{|l|}{022}\\ \hline
BP10 & 76.69 & 154.579 & 163.045 & &$\checked$& & 3.92071 & 0.0096 & 0.028141\\ \hline
BP11 & 98.88 & 155.037 & 155.438 & &&& 1.18124 & -0.0628 & 0.0027369\\ \hline
BP12 & 58.31 & 171.148 & 172.96 &$\checked$&$\checked$&& 0.540354 & 0.00762 & 0.0064099\\ \hline
\multicolumn{10}{|l|}{012}\\ \hline
BP13 & 99.65 & 138.484 & 181.321 & &$\checked$&& 2.46301 & 0.0532 & 0.001255\\ \hline
\textbf{BP14} & 71.03 & 165.604 & 175.971 & $\checked$&$\checked$&$\checked$& 0.339292 & 0.00596 & 0.11841\\ \hline
\textbf{BP15} & 71.03 & 217.656 & 218.738 & $\checked$&$\checked$&$\checked$& 0.766549 & 0.00214 & 0.12225\\ \hline
\multicolumn{10}{|l|}{011}\\ \hline
\textbf{BP16} & 71.33 & 203.796 & 229.092 & $\checked$&$\checked$&$\checked$& 1.03044 & -0.00122 & 0.12214\\ \hline
%
\multicolumn{10}{|l|}{002}\\ \hline
BP18 & 147 & 194.647 & 197.403 &&&& 0.387 & -0.018 & 0.0017718\\ \hline
BP19 & 165.8 & 190.082 & 195.999 &&&& 2.7675 & -0.004 & 0.0028405\\ \hline
BP20 & 191.8 & 198.376 & 199.721 & &&& 1.5075 & 0.008 & 0.008494\\ \hline
\multicolumn{10}{|l|}{001}\\ \hline
\textbf{BP21} & 57.475 & 288.031 & 299.536 & $\checked$&$\checked$&$\checked$& 0.929911 & 0.00192 & 0.11946\\ \hline
\textbf{BP22} & 71.42 & 247.224 & 258.382 &  $\checked$&$\checked$&$\checked$& 1.04301 & -0.00406 & 0.12428\\ \hline
\end{tabular}

\caption{ In all benchmarks $M_h = 125.1$ GeV. Bold font denotes BP with 100\% DM relic density. Accessibility categories are also shown (see text for details).
  Note that BP5 and BP17 were excluded by the updated  XENON1T limits \cite{Aprile:2018dbl}. \label{tab:bench}
}
\end{center}
\end{table}

In our selection, BP1 is an example for a relatively small mass splitting, that forces the intermediate gauge boson to be off-shell. Other benchmark points, as e.g. BP9, allow for on-shell decays in both channels. However, these force the corresponding dark scalar masses to be $\mathcal{O}\lb  150\,\GeV\rb$, therefore leading to smaller production cross sections.

\begin{figure}[ht!]
\begin{center}
  \includegraphics[width=0.75\textwidth]{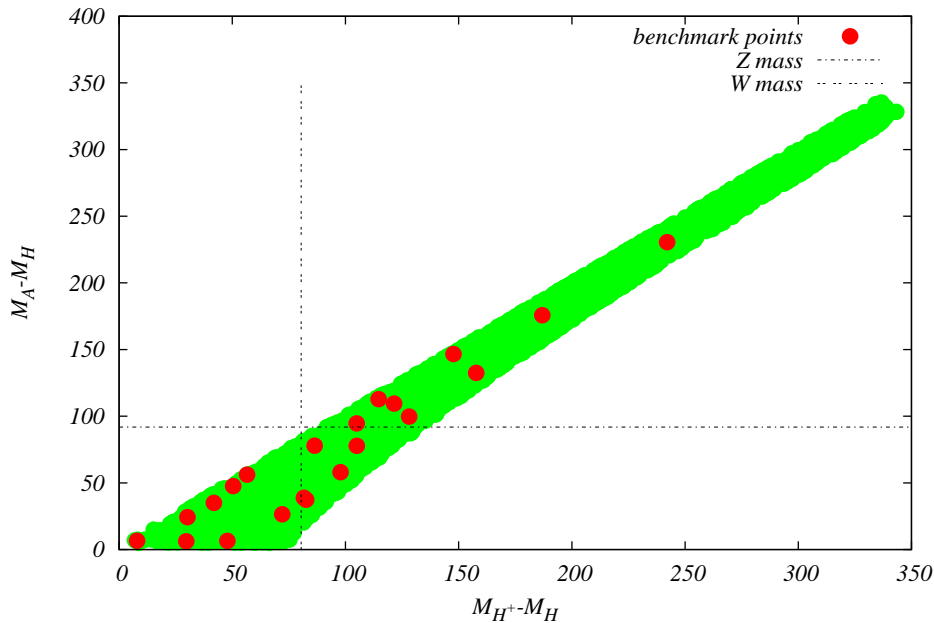}
\end{center}  
\caption{Distribution of benchmark candidate points (green) in the $(M_{H^+}-M_H; M_A-M_H)$ plane, after all constraints are taken into account, as well as selected benchmark points (red) in the same plane. The dashed lines indicate the electroweak gauge boson masses that distinguish between on- and off-shell decays of dark scalars. The relatively narrow band stems mainly from electroweak precision constraints. 
\label{fig:benchmarks}}
\end{figure}

Figure \ref{fig:benchmarks} shows the initial benchmark candidates, that obey all current constraints, in the $(M_{H^+}-M_H; M_A-M_H)$ plane. All points form a narrow band corresponding to $M_A \lesssim M_{H^\pm}$. Our chosen benchmark points, also indicated in Fig.~\ref{fig:benchmarks} (red points) cover mass gaps up to about 250\,GeV only, due to the required minimal cross section (see point 2 above).  Notice that for most selected benchmark points the DM candidate is relatively light, with a mass below 80\,GeV.

\begin{table}[p]
\small
\begin{center}
\begin{tabular}{|l|l|l|l|l|l|l|c|}
\hline 
No. & $M_H$ & $M_A$ & $M_{H^\pm}$ & $\sigma(250)$  & $\sigma(380)$  & $\sigma(500)$ & $BR_{H^+ \to W^+ H}$  \\ 
\hline 
\bf BP1 & 72.77 & 107.803 & 114.639 & 23.7 & 97.8 & 82.6 & $> 0.99$ \\ \hline 
BP2 & 65 & 71.525 & 112.85 & 30.4 & 101 & 83.9 & 0.66 \\ \hline 
BP3 & 67.07 & 73.222 & 96.73 & 108 & 127 & 95.3 & 0.75 \\ \hline 
BP4 & 73.68 & 100.112 & 145.728 & - & 46.7 & 59.9 & 0.92 \\ \hline 
\bf BP6 & 72.14 & 109.548 & 154.761 & - & 33.3 & 53.2 & 0.99 \\ \hline 
BP7 & 76.55 & 134.563 & 174.367 & - & 9.59 & 38.9 & $> 0.99$ \\ \hline 
\bf BP8 & 70.91 & 148.664 & 175.89 & - & 8.16 & 37.8 & $> 0.99$ \\ \hline 
BP9 & 56.78 & 166.22 & 178.24 & - & 6.13 & 36.1 & $> 0.99$ \\ \hline 
BP10 & 76.69 & 154.579 & 163.045 & - & 22.3 & 47.1 & $> 0.99$ \\ \hline 
BP11 & 98.88 & 155.037 & 155.438 & - & 32.4 & 52.7 & 1 \\ \hline 
BP12 & 58.31 & 171.148 & 172.96 & - & 11 & 39.9 & 1 \\ \hline 
BP13 & 99.65 & 138.484 & 181.321 & - & 3.79 & 33.9 & 0.99 \\ \hline 
\bf BP14 & 71.03 & 165.604 & 175.971 & - & 8.09 & 37.7 & $> 0.99$ \\ \hline 
\bf BP15 & 71.03 & 217.656 & 218.738 & - & - & 10.5 & 1 \\ \hline 
\bf BP16 & 71.33 & 203.796 & 229.092 & - & - & 5.64 & $> 0.99$ \\ \hline 
BP18 & 147 & 194.647 & 197.403 & - & - & 23.1 & 1 \\ \hline 
BP19 & 165.8 & 190.082 & 195.999 & - & - & 24 & $> 0.99$ \\ \hline 
BP20 & 191.8 & 198.376 & 199.721 & - & - & 21.6 & 1 \\ \hline 
BP23 & 62.69 & 162.397 & 190.822 & - & - & 27.4 & $> 0.99$ \\ \hline 
\end{tabular}

\caption{ Production cross sections in fb for on-shell charged scalar pair-production, $e^+\,e^-\,\rightarrow\,H^+ H^-$, for the center-of-mass energies considered in this work. We only list benchmark points with at least one non-zero production cross section.  We also display the branching ratio $H^+\,\rightarrow\,W^+\,H$ (the other possible channel, $H^+\,\rightarrow\,W^+ A$, is suppressed for most BPs). The displayed branching ratios were calculated using \texttt{2HDMC}. \label{tab:xshphm} }
\end{center}
\end{table}

\begin{table}[p]
\small
\begin{center}
\begin{tabular}{|l|l|l|l|l|l|l|}
\hline 
No. & $M_H$ & $M_A$ & $M_{H^\pm}$ & $\sigma(250)$  & $\sigma(380)$  & $\sigma(500)$    \\ 
\hline 
\bf BP1 & 72.77 & 107.803 & 114.639 & 77.2 & 65.9 & 45.7  \\ \hline 
BP2 & 65 & 71.525 & 112.85 & 155 & 85.1 & 53.4  \\ \hline 
BP3 & 67.07 & 73.222 & 96.73 & 149 & 83.5 & 52.8  \\ \hline 
BP4 & 73.68 & 100.112 & 145.728 & 89.2 & 69.1 & 46.9  \\ \hline 
\bf BP6 & 72.14 & 109.548 & 154.761 & 75.1 & 65.4 & 45.4  \\ \hline 
BP7 & 76.55 & 134.563 & 174.367 & 31.2 & 52.3 & 40.1  \\ \hline 
\bf BP8 & 70.91 & 148.664 & 175.89 & 20 & 47.5 & 38.1  \\ \hline 
BP9 & 56.78 & 166.22 & 178.24 & 14.1 & 43 & 36  \\ \hline 
BP10 & 76.69 & 154.579 & 163.045 & 9.44 & 43 & 36.2  \\ \hline 
BP11 & 98.88 & 155.037 & 155.438 & - & 35.6 & 33.2  \\ \hline 
BP12 & 58.31 & 171.148 & 172.96 & 9.01 & 40.4 & 34.8  \\ \hline 
BP13 & 99.65 & 138.484 & 181.321 & 5.17 & 42.5 & 36.2  \\ \hline 
\bf BP14 & 71.03 & 165.604 & 175.971 & 5.13 & 39.6 & 34.7  \\ \hline 
\bf BP15 & 71.03 & 217.656 & 218.738 & - & 18.2 & 24.2  \\ \hline 
\bf BP16 & 71.33 & 203.796 & 229.092 & - & 23.3 & 26.9  \\ \hline 
BP18 & 147 & 194.647 & 197.403 & - & 6.14 & 18.7  \\ \hline 
BP19 & 165.8 & 190.082 & 195.999 & - & 3.02 & 16.6  \\ \hline 
BP20 & 191.8 & 198.376 & 199.721 & - & - & 11.3  \\ \hline 
\bf BP21 & 57.475 & 288.031 & 299.536 & - & 2.66 & 12.6  \\ \hline 
\bf BP22 & 71.42 & 247.224 & 258.382 & - & 8.94 & 18.6  \\ \hline 
BP23 & 62.69 & 162.397 & 190.822 & 13.2 & 43.3 & 36.2  \\ \hline 
\end{tabular}

\caption{ Production cross sections in fb for on-shell neutral scalar pair-production, $e^+\,e^-\,\rightarrow\,H A$, for the center-of-mass energies considered in this work. We only list benchmark points with at least one non-zero production cross section. The branching ratio BR$(A\to H Z^{(\star)})\approx 100$\%. \label{tab:xsha} }
\end{center}
\end{table}

Tables \ref{tab:xshphm} and \ref{tab:xsha} show the production cross sections at various center-of-mass energies for all benchmark scenarios.
These indicate promising prospects of detection at future linear colliders. 
Because $A$ is always a lighter particle, neutral channels are usually accessible at lower energies than charged ones. If both channels are accessible, the cross section for charged scalar pair-production is usually larger than for the neutral ones.

\section{Extending to higher mass scales}\label{sec:high}

So far, we have discussed IDM scenarios that are accessible at center-of-mass energies up to 500\,GeV, with a mass range of $M\,\lesssim\,250\,\GeV$ for all dark scalars. However, the parameter space of the IDM largely opens up for higher dark scalar masses, especially as direct detection constraints are less strict and direct collider searches pose less stringent constraints.\footnote{Constraints stemming from the diphoton range exclude certain ranges in the $M_{H^+},\lam_{345}$ plane, cf. e.g. \cite{Ilnicka:2018def}, leading to a lower limit on the dark charged scalar mass.} Therefore, in this section we consider high-mass benchmark points (HP) that can be explored at higher center-or-mass energies, e.g. at 1\,TeV ILC or at the high-energy stages of the CLIC collider, at 1.5\,TeV or 3\,TeV.

As before, all benchmark points have passed theoretical and experimental constraints discussed in section \ref{sec:constraints}. We now consider the nominal collider energies of 1\,TeV, 1.5\,TeV and 3\,TeV. From about 6000 parameter points, we selected 20 benchmark points. The selection was done analogously to the low-energy case, aiming in addition to cover the whole mass range of the dark scalars up to 1\,TeV. Benchmark points are summarized in table \ref{tab:bmhigh}
 {and the production cross sections for the considered IDM scalar pair-production processes in table \ref{tab:bmhigh_cs}}.
The maximum mass splitting between the dark scalars is about 140\,GeV, as the relic density decreases rapidly with the increasing mass difference (see sec.~\ref{sec:exact}).

\begin{table}[p]
\begin{center}
\small
\begin{tabular}{|l|l|l|l|c|c|c|l|l|l|}
\hline 
\multirow{2}{*}{No.} & \multirow{2}{*}{$M_H$} & \multirow{2}{*}{$M_A$} & \multirow{2}{*}{$M_{H^\pm}$} & $Z$ & $W$ & DM & \multirow{2}{*}{$\lambda_2$} & \multirow{2}{*}{$\lambda_{345}$} & \multirow{2}{*}{$\Omega_H h^2$}\\[-1mm] 
 & & & & on-shell &  on-shell & $>$50\%  &  &  & \\ 
\hline 
HP1 & 176 & 291.36 & 311.96 & $\checked$ & $\checked$ &           &  1.4895 & -0.1035 & 0.00072156 \\ \hline 
HP2 & 557 & 562.316 & 565.417 &           &           & $\checked$  &  4.0455 & -0.1385 & 0.072092 \\ \hline 
HP3 & 560 & 616.32 & 633.48 &           &           &           &  3.3795 & -0.0895 & 0.001129 \\ \hline 
HP4 & 571 & 676.534 & 682.54 & $\checked$ & $\checked$ &           &  1.98 & -0.471 & 0.00056347 \\ \hline 
HP5 & 671 & 688.108 & 688.437 &           &           &           &  1.377 & -0.1455 & 0.024471 \\ \hline 
HP6 & 713 & 716.444 & 723.045 &           &           &           &  2.88 & 0.2885 & 0.035152 \\ \hline 
HP7 & 807 & 813.369 & 818.001 &           &           &           &  3.6675 & 0.299 & 0.032393 \\ \hline 
HP8 & 933 & 939.968 & 943.787 &           &           & $\checked$  &  2.9745 & -0.2435 & 0.09639 \\ \hline 
HP9 & 935 & 986.22 & 987.975 &           &           &           &  2.484 & -0.5795 & 0.0027958 \\ \hline 
\bf HP10 & 990 & 992.36 & 998.12 &           &           & $\checked$  &  3.3345 & -0.051 & 0.12478 \\ \hline 
HP11 & 250.5 & 265.49 & 287.226 &           &           &           &  3.90814 & -0.150071 & 0.00535 \\ \hline 
HP12 & 286.05 & 294.617 & 332.457 &           &           &           &  3.29239 & 0.112124 & 0.00277 \\ \hline 
HP13 & 336 & 353.264 & 360.568 &           &           &           &  2.48814 & -0.106372 & 0.00937 \\ \hline 
HP14 & 326.55 & 331.938 & 381.773 &           &           &           &  0.0251327 & -0.0626727 & 0.00356 \\ \hline 
HP15 & 357.6 & 399.998 & 402.568 &           &           &           &  2.06088 & -0.237469 & 0.00346 \\ \hline 
HP16 & 387.75 & 406.118 & 413.464 &           &           &           &  0.816814 & -0.208336 & 0.0116 \\ \hline 
HP17 & 430.95 & 433.226 & 440.624 &           &           &           &  3.00336 & 0.082991 & 0.0327 \\ \hline 
HP18 & 428.25 & 453.979 & 459.696 &           &           &           &  3.87044 & -0.281168 & 0.00858 \\ \hline 
HP19 & 467.85 & 488.604 & 492.329 &           &           &           &  4.12177 & -0.252036 & 0.0139 \\ \hline 
HP20 & 505.2 & 516.58 & 543.794 &           &           &           &  2.53841 & -0.354 & 0.00887 \\ \hline 
\end{tabular}

\caption{High-mass benchmark points (HPs) accessible at  {linear} colliders with $\mathcal{O}\lb\TeV\rb$ center-of-mass energies. $M_h=125.1\,\GeV$ for all points. HP10 provides exact relic density. \label{tab:bmhigh} }
\end{center}
\end{table}

\begin{table}[p]
\begin{center}
\small
\begin{tabular}{|l|l|l|l|c|c|c|c|c|c|}
\hline 
No. & $M_H$ & $M_A$ & $M_{H^\pm}$ & \multicolumn{3}{c|}{$\sigma(e^+e^-\to H^+H^-)$}& \multicolumn{3}{c|}{$\sigma(e^+e^-\to AH)$}  \\ 
\cline{5-10} 
  & [GeV] & [GeV] & [GeV] & 1\,TeV & 1.5\,TeV & 3\,TeV & 1\,TeV & 1.5\,TeV & 3\,TeV   \\ 
\hline 
HP1 & 176 & 291.36 & 311.96 & 13 & 9.9 & 3.4  &  8.6 & 5.1 & 1.6 \\ \hline 
HP2 & 557 & 562.316 & 565.417 & - & 3.1 & 2.6  &  - & 1.4 & 1.1 \\ \hline 
HP3 & 560 & 616.32 & 633.48 & - & 1.6 & 2.4  &  - & 1.1 & 1.1 \\ \hline 
HP4 & 571 & 676.534 & 682.54 & - & 0.68 & 2.2  &  - & 0.77 & 1.1 \\ \hline 
HP5 & 671 & 688.108 & 688.437 & - & 0.59 & 2.2  &  - & 0.32 & 0.98 \\ \hline 
HP6 & 713 & 716.444 & 723.045 & - & 0.16 & 2.1  &  - & 0.11 & 0.93 \\ \hline 
HP7 & 807 & 813.369 & 818.001 & - & - & 1.8  &  - & - & 0.79 \\ \hline 
HP8 & 933 & 939.968 & 943.787 & - & - & 1.4  &  - & - & 0.6 \\ \hline 
HP9 & 935 & 986.22 & 987.975 & - & - & 1.2  &  - & - & 0.57 \\ \hline 
\bf HP10 & 990 & 992.36 & 998.12 & - & - & 1.2  &  - & - & 0.52 \\ \hline 
HP11 & 250.5 & 265.49 & 287.226 & 15 & 11 & 3.5  &  7.8 & 4.9 & 1.6 \\ \hline 
HP12 & 286.05 & 294.617 & 332.457 & 11 & 9.4 & 3.3  &  6.5 & 4.6 & 1.5 \\ \hline 
HP13 & 336 & 353.264 & 360.568 & 8.5 & 8.6 & 3.3  &  4.3 & 3.9 & 1.4 \\ \hline 
HP14 & 326.55 & 331.938 & 381.773 & 6.7 & 8 & 3.2  &  4.9 & 4.1 & 1.5 \\ \hline 
HP15 & 357.6 & 399.998 & 402.568 & 5 & 7.5 & 3.1  &  3 & 3.5 & 1.4 \\ \hline 
HP16 & 387.75 & 406.118 & 413.464 & 4.2 & 7.2 & 3.1  &  2.4 & 3.3 & 1.4 \\ \hline 
HP17 & 430.95 & 433.226 & 440.624 & 2.4 & 6.4 & 3  &  1.3 & 2.9 & 1.3 \\ \hline 
HP18 & 428.25 & 453.979 & 459.696 & 1.3 & 5.9 & 3  &  1 & 2.8 & 1.3 \\ \hline 
HP19 & 467.85 & 488.604 & 492.329 & 0.09 & 5 & 2.8  &  0.21 & 2.4 & 1.3 \\ \hline 
HP20 & 505.2 & 516.58 & 543.794 & - & 3.7 & 2.7  &  - & 2 & 1.2 \\ \hline 
\end{tabular}

\caption{Production cross sections in \fb~for high-mass benchmark points at 1\,TeV, 1.5\,TeV and 3\,TeV, for the production processes considered here.\label{tab:bmhigh_cs}}
\end{center}
\end{table}

 {
\section{Prospects for future dark matter and LHC experiments }\label{sec:lhc}

Experiments at future $e^+ e^-$ colliders will be able to probe the described benchmark
scenarios on the time scales of ten to twenty years.
By that time, the sensitivity of direct DM search experiments will have improved significantly, and
much larger samples of data will also be collected by the LHC experiments.
It is therefore an important question whether the benchmarks presented here are already accessible at the LHC or via dark matter direct detection. 

As already described above, significant constraints of the IDM scenarios are set
by the recent  XENON1T measurement \cite{Aprile:2018dbl}.
The experiment continues to collect data and it is expected to improve its
sensitivity by a factor of about 3 \cite{Aprile:2015uzo}.
In this range, {most} of the points {providing dominant contribution to the
DM density in the Universe} (see tables \ref{tab:bench} and \ref{tab:bmhigh})
can be probed.
Four of these scenarios (BP6, BP16, BP21 and HP10) will remain inaccessible
at XENON1T, but can be probed at XENONnT, when the sensitivity will be enhanced
by another order of magnitude.
Only one scenario, BP2, will not be probed after the future XENON detector upgrade.

Scenarios resulting in subdominant contribution to the relic density are less
constrained by direct search experiments.
Still, half of the presented benchmark points can be probed with the full XENON1T
data  and only four scenarios (BP18, BP19, BP20 and HP3) remain inaccessible at
XENONnT.
This shows that direct DM search experiments will continue to set important
constraints on the proposed class of models and can result in the need to
redefine the set of benchmark points in the future.
On the other hand, if an excess in direct detection is observed
in the XENON detector, the IDM scenarios will provide a perfect test framework
for interpretation of these results. 
Direct searches will only indicate the possible mass and coupling range
for the lightest IDM scalar.
Other parameters of the model need to be constrained in collider experiments.

Searches for pair production of IDM scalars have also been considered for current and future runs of the  LHC.
Many different final states can be considered, including mono-jet, mono-Z,
mono-Higgs and Vector-Boson-Fusion + missing transverse energy signatures \cite{Poulose:2016lvz,Datta:2016nfz,Belyaev:2016lok,Dutta:2017lny,Wan:2018eaz}. Especially multi-lepton final states \cite{Datta:2016nfz} are promising\footnote{A discussion on regions accessible at Run I can e.g. be found in \cite{Belanger:2015kga}.}. In addition, also multi-jet final states \cite{Poulose:2016lvz,Dutta:2017lny} and combinations \cite{Wan:2018eaz} have been considered. For example, in \cite{Datta:2016nfz,Dutta:2017lny,Wan:2018eaz} the authors present phenomenological studies that render scenarios with dark masses $\lesssim\,300\,\GeV$ accessible at the HL-LHC.

Regarding other channels,
the analysis of mono-jet signature considered in \cite{Belyaev:2018ext,Belyaev:2016lok}
indicates that LHC has limited sensitivity to probe the IDM.
Additional assumptions are needed to set the scalar mass limits and they range only
up to about 200 GeV for HL-LHC.
Also for the di-jet plus missing transverse energy signature, as studied
e.g. in \cite{Poulose:2016lvz}, LHC sensitivity is significantly affected
by the large background, which can not be sufficiently suppressed even with
strong kinematic selection.
For the benchmark scenarios considered, the maximum signal significance
at an integrated luminosity of 3000 fb$^{-1}$ was about 2$\sigma$. However, we want to state that these analyses have not been on the same level as the current study, as they did not make use of more advanced analysis techniques. In general, production cross sections can be in the $\mathcal{O}\lb 500-700\fb\rb$ range for a center-of-mass energy of 13 \TeV~ \cite{trtalk}. Without detailed analyses, projections of reachability are however difficult to make. We therefore strongly encourage the experimental collaborations to investigate the benchmark points presented here at current and future LHC runs.

}

\section{Conclusions}\label{sec:conclusions}
In this paper we have revisited and updated the available parameter space of the Inert Doublet Model. The model features an exact $Z_2$ symmetry which results in dark scalars that do not interact with SM fermions, and the lightest neutral scalar can serve as a promising dark matter particle.
We took into account most recent experimental constraints from relic density and direct dark matter searches, including the latest XENON1T 2018 results, as well as collider bounds and theoretical constraints. Based on these updated results, we have provided benchmark scenarios accessible at the initial stages of future linear $e^+e^-$ colliders (250 and 500 GeV ILC and 380 GeV CLIC) as well as benchmarks that can be tested at high-energy stages (1 TeV ILC and 1.5 and 3 TeV CLIC).
In doing so  we pursued the philosophy of covering the widest range of parameters and experimental signatures. We provide predictions of production cross sections at these energies, and supplement these with information about the branching fractions of the relevant decay modes. We encourage the LC groups to make use of these benchmark scenarios. Although the benchmarks have been defined with $e^+e^-$ physics in mind, we strongly encourage our LHC experimental colleagues to consider these scenarios in the analysis of the current and upcoming LHC data.

\section*{Acknowledgements}
The authors thank J\"urgen Reuter and Wolfgang Kilian for useful comments regarding \texttt{Whizard}
and the \texttt{Sarah}/\texttt{Spheno} interface.
This research was supported in parts by the National Science Centre,
Poland, the HARMONIA project under contract UMO-2015/18/M/ST2/00518
(2016-2019) and OPUS project under contract UMO-2017/25/B/ST2/00496
(2018-2021) as well as COST Action CA 15180. The work of TR was partially supported by
the National Science Foundation under
Grant No. 1519045, by Michigan State University through computational resources
provided by the Institute for Cyber-Enabled Research, and by grant K 125105 of
the National Research, Development and Innovation Fund in Hungary.
The work of WK was partially supported by the German Research Foundation (DFG) under grants
number STO 876/4-1 and STO 876/2-2. JK and WK thank Gudrid Moortgat-Pick for her hospitality and the DFG for support 
through the SFB~676 ``Particles, Strings and the Early Universe'' 
during the initial stage of this project.

\bibliographystyle{jhep}
\bibliography{lit}

\end{document}